\documentclass[final,5p,times,twocolumn]{elsarticle}

\usepackage{graphics,graphicx,dcolumn,bm,fleqn,epic,eepic,float,ulem}
\usepackage{amssymb,amsmath}
\usepackage{multirow,rotate}
\usepackage{color}
\usepackage[displaymath]{lineno}
\usepackage[english]{babel}
\usepackage{lscape}
\usepackage{rotating}
\usepackage{pdflscape}
\usepackage{tabularx}
\usepackage{colortbl}

\definecolor{red}{rgb}{1,0,0}
\definecolor{green}{rgb}{0,1,0}
\definecolor{blue}{rgb}{0,0,1}
\definecolor{dark_grey}{gray}{0.3}
\definecolor{ligth_grey}{gray}{0.7}

\newcommand{\no}{NO$_2$}

\newcommand{\pa}{PM$_{10}$}
\newcommand\panel[1]{\multicolumn{1}%
{>{\columncolor{ligth_grey}}#1}}

\journal{Science of the SET Environment}
\begin{document}
%%%%%%%%%%%%%%%%%%%%%%%%%%%%%%%%%%%%%%%%%%%%%%%%%%%%%%%%%%%%%%%%%%%%%%%%%

\begin{frontmatter}

%\title{\pa\ forecast using synoptic and local scale meteorological data}
\title{Daily pollution forecast using optimal meteorological data at synoptic and local scales}

\author{Ana Russo$^{a}$}
\author{Pedro G.~Lind$^{c,d}$}
\author{Frank Raischel$^{a,b}$}
\author{Ricardo Trigo$^{a}$}
\author{Manuel Mendes$^{e}$}
%\author{Isabel F. Trigo$^{b}$}

\address{$^a$Center for Geophysics , IDL, 
         University of Lisbon
         1749-016 Lisboa, Portugal}
\address{$^b$Center for Theoretical and Computational Physics, 
         University of Lisbon, Av.~Prof.~Gama Pinto 2, 
         1649-003 Lisbon, Portugal}
\address{$^c$TWIST - Turbulence, Wind energy and Stochastics,
         Institute of Physics, Carl-von-Ossietzky University of 
         Oldenburg, DE-26111 Oldenburg, Germany}
\address{$^d$ForWind - Center for Wind Energy Research, Institute of Physics,
         Carl-von-Ossietzky University of Oldenburg, DE-26111 Oldenburg, 
         Germany}
\address{$^e$Instituto de Meteorologia, Rua C-Aeroporto,1749-077 Lisbon, Portugal}

%\begin{linenumbers}
\begin{abstract} 
%Air pollution is a complex mixture of toxic components that might have serious impacts 
%on human health and quality of life, specially for sensitive groups, such as children and 
%people with heart and respiratory insufficiencies. 
%Air pollution is determined by the combination of different mechanisms, namely, emissions, 
%geographical constrains, meteorological and chemical processes. 
%The relative importance of such factors is influenced by their interaction on diverse scales 
%of atmospheric motion where they constitute a highly non-linear and fluctuative phenomenon 
%which is not easy to estimate in advance. 
%Thus, forecasting models emerge as an effective approach in order to identify and 
%predict episodes of high pollution levels.
We present a simple framework to easily pre-select the most essential data for 
accurately forecasting the concentration of the pollutant \pa,
based on pollutants observations for the years 2002 until 2006 in the metropolitan
region of Lisbon, Portugal.
Starting from a broad panoply of different data sets collected at several meteorological 
stations, we apply a forward stepwise regression procedure that enables us not only
to identify the most important variables for forecasting the pollutant but also
to rank them in order of importance.
We argue the importance of this variable ranking, showing that the ranking is very
sensitive to the urban spot where measurements are taken.
Having this pre-selection, we then present the potential of linear and non-linear 
neural network models when applied to the concentration of pollutant \pa.
Similarly to previous studies for other pollutants, our validation results show that
non-linear models in average perform as well or worse as linear models for \pa. 
Finally, we also address the influence of Circulation Weather Types, characterizing
synoptic scale circulation patterns and the concentration of pollutants.
%Here, we present the potential of linear and non-linear neural network models when 
%applied to the concentration of pollutant \pa, measured at the urban area of Lisbon, 
%Portugal. The interaction between meteorological variables and pollutants is studied 
%locally, based on pollutants observations for the years 2002 until  2006. 
%Furthermore, the choice of the 
%best predictors for each monitoring station is performed independently and is 
%determined by stepwise regression 
%analysis. 
%Similarly to previous studies for other pollutants, our validation results show that
%non-linear models in average perform as well or worse as linear models for \pa. 
%Finally, we briefly discuss other possible applications of such optimized neural network 
%models. 
\end{abstract}

%%%%PACS e Keywords
%\begin{keyword}
%Pollutants \sep
%Neural Networks \sep
%Circulation weather types\sep
%Meteorology \sep
%Environmental Research 
%%\PACS[2010] 92.60.Sz \sep  % Air pollution
%%            02.50.Ga \sep  %Markov processes 
%%            02.50.Ey \sep  %Stochastic processes
%%           92.70.Gt       %Climate dynamics 
%\end{keyword}

%\end{linenumbers}
\end{frontmatter}

%%%%%%%%%%%
%\begin{linenumbers}
\section{Introduction}
\label{sec:intro}

Air pollution is a global threat to public health and to the environment, although its effects are generally strongest in urban areas \citep{kolehmainen2001, EEA, EEA1, EEA2}.
Urban air pollution is a complex mixture of toxic components, which may induce acute 
and chronic responses from sensitive groups, such as children and people with previous heart 
and respiratory insufficiencies \citep{kolehmainen2001,Wong,diaz2004}. 
Therefore, forecasting the temporal evolution of air pollution concentrations in urban 
locations emerges as a priority for guaranteeing life quality in urban areas \citep{kolehmainen2001, EEA1, EEA2}.

Modelling air pollution allows to describe the causal relationship between emissions, 
meteorology, atmospheric concentrations, deposition, and other factors, including  the determination of the effectiveness 
of remediation strategies, and the simulation of future scenarios. Different types of approaches have been applied to characterize 
and forecast the dispersion of air pollutants, 
from the most simple approaches, such as %(e.g., 
box models \citep{middleton1997}, Gaussian plume models \citep{reich1999}, persistence 
and regression models \citep{shi1999}, to the most complex model systems, namely % (e.g., 
UAM-Urban Airshed Model \citep{Morris}, ROM-Regional Oxidant Model \citep{Davis}, 
CHIMERE \citep{Monteiro}, CMAQ-Community Multiscale Air Quality 
Model \citep{luecken2006,sokhi2006,arasa}.

Simpler models are used often as they can provide a fast overview. However they rely on significant simplifying 
assumptions and usually do not describe the complex processes and interactions that control 
the transport and chemical behavior of pollutants in the atmosphere \citep{luecken2006}. 

For detailed characterization of atmospheric pollution more sophisticated models are needed, such as 
dispersion models which are driven by the objective quantification of chemical reactions and the 
physical transport of pollutants. In the last decades, significant progress has been
made in air-quality dispersion models \citep{arasa}. 
However, being highly non-linear, they require 
large amounts of accurate input data and are computationally expensive \citep{dutot}.

Statistical models, such as Artificial Neural Networks (NN), 
can constitute a promising alternative to deterministic models \citep{YiPrybutok1996, 
Cobourn2000}, namely in what concerns air pollution problems \citep{dutot, binbhu2012,GardnerDorling2000a, 
Hooyberghs2005, Papanastasiou, nejadkoorki2012}. 
These models are usually regarded as a good compromise between simplicity and effectiveness, being capable of modeling the effect of non-linearities and fluctuations. 

Although NN models may involve greater uncertainty than more complex models, the input 
data requirements are less strict. 
Several NN models were already tested, mostly for forecasting hourly averages 
\citep{kolehmainen2001, perez2000, kukkonen2003} or daily maxima \citep{perez2002}
 of air pollutants.
Some authors compared the potential of different approaches when applied to different 
pollutants and prediction time lags \citep{YiPrybutok1996, GardnerDorling2000a, 
Hooyberghs2005, kukkonen2003}. Other authors have proven better forecasting results 
of NN over multiple linear regression (MLR) \citep{perez2000, kukkonen2003, 
Agirre-Basurko2006}. 
More recently, some of us \cite{russo} showed that, combining NN models and stochastic 
data analysis, allows to diminuish the requirement of large training data sets often
appearing when constructing a NN model.

Despite these improvements forecasting NN models still present some caveats that need to be properly addressed.
First, the time-lag in which air pollution prediction is performed should be as
large as possible for enabling effectiveness of alert procedures in urban 
centers.
Although hourly NN modeling has been frequently and successfully applied in 
air pollution studies, 
modeling daily concentration is more adequate to enable useful information
to citizens 
%modeling daily concentration averages have  rarely been attempted 
\citep{Hooyberghs2005}.

Second, the construction of the best NN structure and the choice of input parameters constitutes another challenge for modelers \citep{Hooyberghs2005}.
Theoretically, any set of input data can be fed into any NN architecture for training and evaluation. However, the number of possible predictors and the number of ways they can be presented is too diverse to test all possible combinations. Here, we decided to use two of the most common architectures used ir air quality modelling, a linear and non-linear NN. The linear NN model is based on a simple one-layer structure which produces the same results as a linear regression model \citep{Weisberg1985} and the non-linear NN models are based on a feed-forward configuration of the multilayer perceptron that has been used by several authors \citep{Hooyberghs2005, Papanastasiou, nejadkoorki2012}. Regarding the choice of the best input parameters, we argue that besides the intrinsic parameters describing air pollution, the interaction between pollutants and weather patterns
should have the potential to significantly improve  air quality forecasts.

Third, while  several studies have been published establishing links between
synoptic scale circulation patterns, usually named Circulation Weather Types (CWT) 
and air pollution \citep{DayanLevy2002, demuzere,  carvalho2010, Saavedra2012},
the majority of the research focused on individual meteorological 
variables and non-automated procedures of variables’ selection.
Weather is one of the factors that conditions air quality 
\citep{DayanLevy2002, Baumbach1996}, constraining the atmospheric processes that 
are associated to the occurrence of pollution episodes, namely, the processes of 
dilution, transformation, transport and removal of pollutants\citep{Baumbach1996}.
The relative importance of weather and climate for predicting the state
of air quality has been investigated extensively over the last few 
decades\citep{DayanLevy2002, demuzere,  carvalho2010, Pearce2011, Saavedra2012, no2}. 
Those studies revealed that certain weather parameters are relevant to model air 
pollutant concentrations, particularly, the temperature, wind speed and direction, relative 
humidity, cloud cover, dew point temperature, sea level pressure, precipitation 
and mixing layer height \citep{Hooyberghs2005, demuzere}. 
However, the majority of the research focused on individual meteorological 
variables and non-automated procedures of variables’ selection. 
%Moreover, several studies have been published establishing links between
%synoptic scale circulation patterns, usually named Circulation Weather Types (CWT) and 
%air pollution \citep{DayanLevy2002, demuzere,  carvalho2010, Saavedra2012};
%and linking a particular air mass to dispersion conditions and also to the
%mesoscale and local meteorological behaviour \citep{DayanLevy2004}. 
%Thus, the relationships between air pollution and meteorological variables or atmospheric synoptic patterns represent an important research area.
%These prevailing CWT can be determined at regional scale and therefore it is
%possible to evaluate the relative importance of each CWT in pollutant concentration 
%forecasts. 
A specific application to pollution in the Iberian Peninsula was developed by Saavedra et al. (2012), who presented a very detailed description of the relationship between synoptic pressure patterns and high-ozone incidents in northwest Galicia. Nevertheless, there are to the best of our knowledge no studies in the literature focusing on the application over the Iberian Peninsula of objective automatic classification procedures of CWT as a predictor for air quality modelling.
Bringing the insight of such previous studies and considering CWTs
as possible input parameters for NN models could provide better 
forecasts, particularly for large time-lags.
The proposed approach should also allow to ascertain how strongly do meteorological variables and CWT influence
the concentration of pollutants.

In this paper, we address all these three issues, aiming on daily forecast,
introducing a simple framework for automatically rank the set of variables used
as input variables for training the NN model and, in particular, 
addressing the influence of CWT in air pollution evolution.
More exactly, 
we study relations between weather and air pollution through a circulation-to-environment approach
based on the analysis of the existence of links between meteorological parameters and daily air quality 
measurements. 
This study focuses on the development of air quality models within the greater 
urban area of Lisbon, Portugal, based on an approach that is able 
to capture the temporal evolution of air quality and to produce corresponding
forecasts. 
We choose to address one single pollutant belonging to the group
of particulate matter (PM), namely \pa.
Atmospheric PM comprehends in general 
%exceptionally 
small particules of solid or liquid matter 
in suspension or associated with atmospheric gases. Usually this mixture
of air and PM is called atmospheric aerosol. Typically, these cover a
wide range of sizes and chemical characteristics, including PAHs, acid aerosols and
diesel particulates \citep{EEAr}. \pa\ include respirable particulate matter sized 10 $\mu g$ or less, 
which poses a major health risk \citep{EEAr}.
Although pollutants' emissions have decreased over the last two decades, 
this did not lead to a corresponding reduction of concentrations of \pa\ 
throughout Europe \citep{EEA}, which influenced 
the choice of this pollutant as target of this study. 
Additionally, evidence has accumulated during the last years that there is a direct association between daily variations in the concentrations of
airborne particles and a range of health indicators, which include daily deaths, admissions to
hospital for the treatment of both respiratory and cardiovascular diseases and symptoms amongst
patients suffering from asthma \citep{Wong,diaz2004,EEAr}. 
For the particular case of \pa\, since 2005, European Union imposes a 
limiting value of $50 \mu g$ per cubic meter of air \cite{eu}, with a number
of exceeding values not more than $35$ per year.
Lisbon is located closely to the Atlantic ocean where most of the 
moisture affecting western Iberia arrives \citep{Gimeno2010}, particularly in winter 
months \citep{Trigo2002}. Despite this impact of the ocean that mitigates
the effects of aerosols and pollution, Lisbon has been affected by 
several high pollution episodes in the last two decades, exceeding repeatedly
the legal limits imposed for \pa\ \citep{APA1}.
Therefore, and facing such restrictive rules and the exceeding events that occurred on the last years, a good \pa\ prediction procedure within a sufficiently
large time-lag to prevent the occurrence of exceeding concentrations is needed.

To perform a systematic study, we apply both linear and non-linear NN
models to predict \pa\ daily average concentrations based on 
air pollution and weather historical information.

Comrie \citep{Comrie} and Cobourn et 
al. \citep{Cobourn2000} have performed comparison studies between NN 
and regression models to forecast ozone concentrations, both showing 
that NN outcomes are only equal or slightly better than regression models 
for ozone prediction. In contrast, Gardner and Dorling (2000) \citep{GardnerDorling2000b} showed that there is a significant increase in performance
when using non-linear models, although regression models allow to easily undestand and interpret results in terms of the
physical mechanisms between meteorological and air quality variables.
%%%%%%%%%%%%%%%%%%%%%%%%%%%%%%%%%%%%%%%%%%%%%%%%%%%%%%%%%%%%%%%%%%%%
\begin{figure*}[t]
  \centering
  \includegraphics[width=0.74\textwidth]{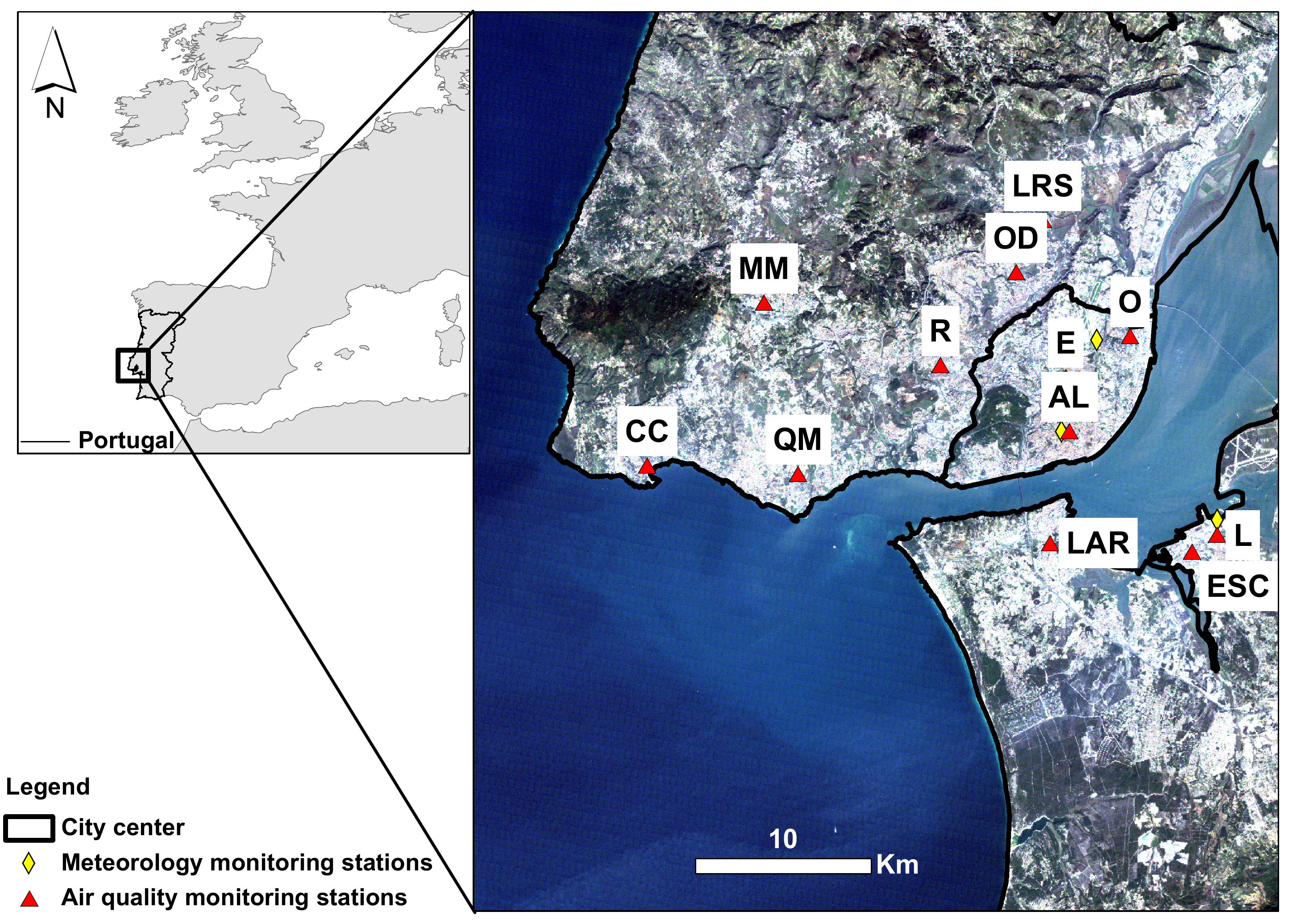}
  \caption{\protect 
           Air quality and meteorological monitoring stations in 
           the region of Lisbon (Portugal). 
           Each data set is extracted within the period between 2002 
           and 2006, with a sampling frequency of $1$ day$^{-1}$.} 
\label{fig1}
\end{figure*}
%%%%%%%%%%%%%%%%%%%%%%%%%%%%%%%%%%%%%%%%%%%%%%%%%%%%%%%%%%%%%%%%%%%%

Three important components will be incorporated:
\begin{enumerate}
\item An important aspect addressed by us is periodicity.
Since the factors mainly contributing to air pollution concentration are 
connected with source activity and periodic variations in nature, it is 
normal to expect periodic components in air quality time series 
\citep{kolehmainen2001}. Hence, following a similar approach to the study presented by Kolehmainen et al. (2001) \citep{kolehmainen2001}, two modeling approaches are possible.
One is to model the original and complete signal. Another is to model the residual 
component after the removal of a periodic component from the complete 
signal.
Here, we address the relative importance of the weekly periodic component. The weekly component is mainly affected by traffic and weekly business and industrial flutuations.
The forecasting capabilities of the different approaches are compared.

\item Furthermore, we also focus on the advantages of the application of an 
automated procedure for the selection of variables \cite{wilks2006}, 
recently used by us \cite{russo,no2}: 
the use of an automated procedure prior to NN 
modeling allows for substantial reduction in the number of input variables, 
which enables also to improve the quality and robustness of pollutant 
concentration forecasts. These are crucial properties when linking the 
forecast to alert systems. 

\item Finally, we use data from several monitoring stations in Lisbon. 
Therefore, our predictions will allow us to define air pollution episode 
alerts with spatial variability, instead of a unique value representing 
the entire region of the urban center.
All in all, even though performance indicators resulting from modeling 
daily concentration averages will be expectantly lower than those attained 
for hourly predictions, the methodological approach here presented can be 
relevant for daily surveillance and alert systems in the Lisbon area.
\end{enumerate}

We start in Sec.~\ref{sec:data} by describing the empirical data, comprising
the different data sets in the city of Lisbon, Portugal (see Fig.~\ref{fig1}). 
In Sec.~\ref{sec:methods} we briefly describe NNs models as well as the main 
points of the circulation-to-environmental approach used and in 
Sec.~\ref{sec:results} the results are discussed in the light of predictive 
power measures and independent validation of our model is provided. 
Section \ref{sec:conclusions} concludes the paper.

%%%%%%%%%%%%%%%%%%%%%%%%%%%%%%%%%%%%%%%%%%%%%%%%%%%%%%%%%%%%%%%%%%%%%%%%%%%%%
\section{Data}
\label{sec:data}

\subsection{Target data}
\label{sec:targetdata}

We consider daily measurements of \pa\ concentrations measured by
twelve monitoring stations in the agglomeration of Lisbon, between 2002 and 
2006 (see bullets in Fig.~\ref{fig1}).

The Lisbon agglomeration is covered by a conventional air quality 
monitoring network composed by traffic, industrial and background monitoring 
stations which record the atmospheric concentrations of major pollutants,
such as gases like NO$_2$, NO and CO and aerosols like PM$_{10}$.
This monitoring network for air quality is complemented by three 
meteorological monitoring stations (see diamonds in Fig.~\ref{fig1}), 
located near the stations of Avenida da Liberdade (AL), Lavradio (L) and 
Olivais (O).
%%%%%%%%%%%Tabela com as variaveis utilizadas por cada modelo%%%%%%%%%%%%%%
\begin{table}[t]
\begin{center}
%\begin{ruledtabular}
\small\begin{tabular}{ l }
\hline\\%[10pt]
{\bf Variables }\textit{(Lag=1 day)}\\[10pt]
\hline\\
Mean concentration of \no, NO, CO, \pa \\[3pt]
Maximum concentration of \pa\ (\pa\ m)\\[3pt]
Concentration of \pa\ at 0h UTC \\[3pt]
Daily circulation weather type (CWT) \\[3pt]
Boundary layers heights: \\[3pt]
\ \ \ (BLH5) at 03:00 UTC \\[3pt]
\ \ \ (BLH7) at 09:00 UTC \\[3pt]
\ \ \ (BLH11) at 21:00 UTC \\[3pt]
Daily maximum temperature (Tmax)\\[3pt]
Daily mean wind direction ($V_d$) and intensity ($V_i$)\\[3pt]
Daily mean humidity (Hum) and radiance (Rad)\\[3pt]
\hline
\end{tabular}
%\end{ruledtabular}
\end{center}
  \caption{\protect 
           Input parameters used for training the NN (see text).}       
           \label{fig_tabela_var}
\end{table}
%%%%%%%%%%%%%%%%%%%%%%%%%%%%%%%%%%%%%%%%%%%%%%%%%%%%%%%%%%%%%%%%%%%

\begin{center}
\begin{figure}[H]
  \centering
  \includegraphics[width=0.45\textwidth]{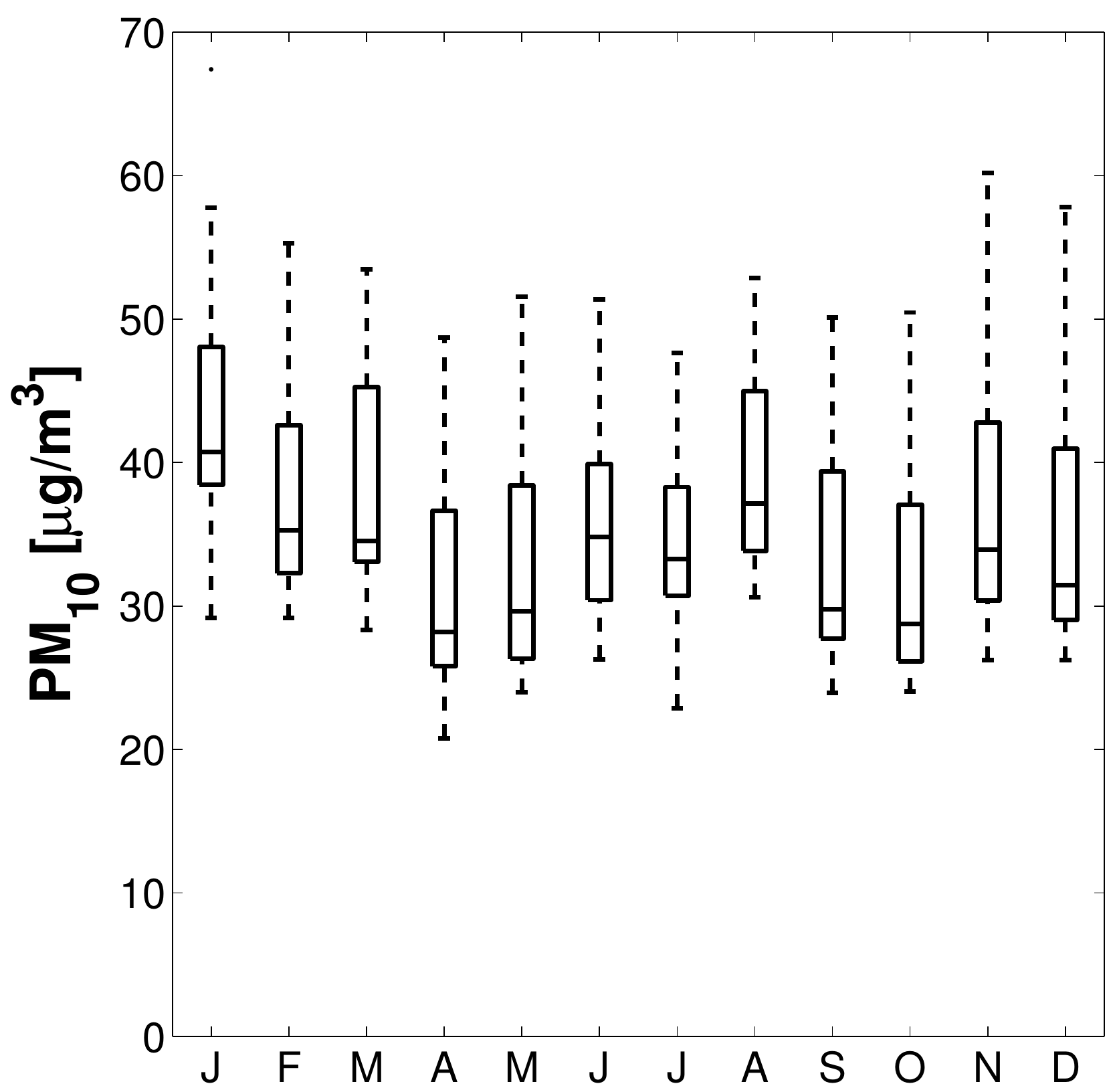}\\[3pt]
	\includegraphics[width=0.45\textwidth]{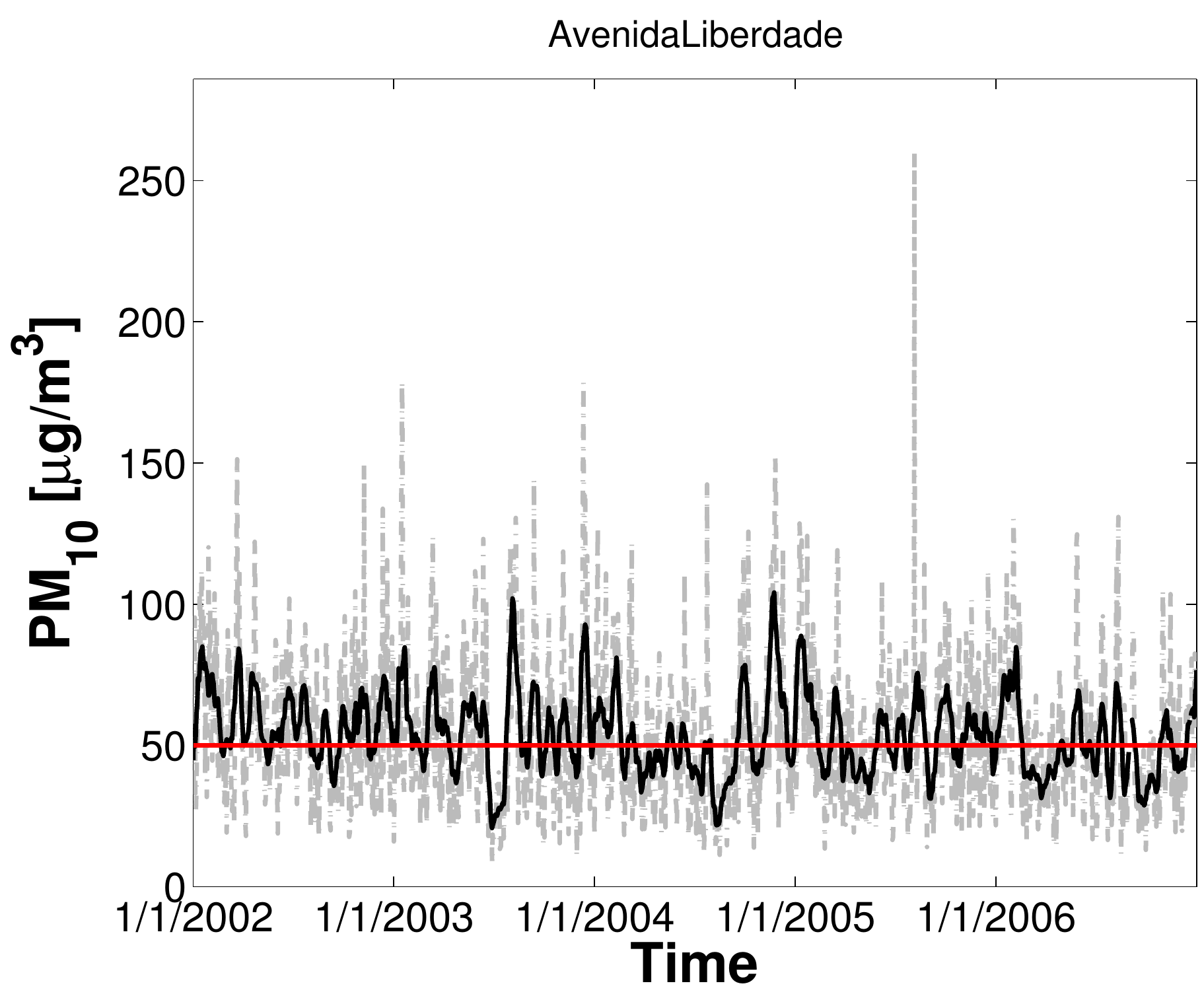}
%  \caption{\protect 
%           Monthly mean distributions of \pa\ concentrations 
%           for the years 2002 till 2006 in Lisbon.}
%\label{fig2}
%\end{figure}
%%%%%%%%%%%%%%%%%%%%%%%%%%%%%%%%%%%%%%%%%%%%%%%%%%%%%%%%%%%%%%%%%%%%
%%%%%%%%%%%%%%%%%%%%%%%%%%%%%%%%%%%%%%%%%%%%%%%%%%%%%%%%%%%%%%%%%%%%
%\begin{figure}[H]
%  \centering
%	(a)	\hspace{8.2cm}(b)
  \caption{\protect 
           (a) Monthly mean distributions of \pa\ concentrations 
           for the years 2002 till 2006 in Lisbon; (b) PM10 concentrations for the period 2002-2006 recorded at
           Avenida da Liberdade (AL) monitoring station. The light grey line 
           represents PM10 daily measures, the black line represents 
           the 7 days moving average and the horizontal line refers to the \pa\ daily legal limit ($50 \mu gm^{-3}$).}
\label{fig2}
\end{figure}

\end{center}

To investigate the presence of seasonal cycles 
a preliminary data analysis is done, yielding the box-plot in Fig.~\ref{fig2} (a), 
showing the monthly distribution of pollutant's concentrations throughout the 
year for the entire studied period (2002-2006) and for all the monitoring stations.
Figure \ref{fig2}(b) supplements the previous box-plot analysis %but just 
just for Avenida da Liberdade (AL) monitoring station: %, and 
no clear annual cycle can be drawn.

Daily legal limits were often exceeded during the 2002-2006 period in all the monitoring stations \citep{APA1, APA2}, but the number of days with exceeding values is specially impressive for Avenida da Liberdade (AL) and Entrecampos (E) stations. It is worth mentioning that, in both stations two types of exceedences occur. On one hand, the daily legal limit ($50 \mu gm^{-3}$) is exceeded, but the number of times that the daily limit can be exceeded per year (35 excedeences/year) is also surpassed \citep{APA1, APA2}.

\subsection{Input data for NN training}
\label{sec:inputdata}

The NN input data sets are shown in Table~\ref{fig_tabela_var} and consist of daily concentration measurements of several 
pollutants besides \pa\ (the target), namely \no, NO, and CO.
Additionally to the pollutant's concentration measured on the previous day and at 0h UTC (Universal Time Coordinated), several meteorological variables measured in the 3 monitoring stations available were considered (Table \ref{fig_tabela_var}).

In order to include information regarding the atmospheric stability and circulation, which is an important factor for the accumulation of pollutants near the surface, two other variables were considered, namely the boundary layer and the daily CWT. The boundary layer height (BLH)
fields were retrieved from the ECMWF 40 years 
reanalysis\cite{web2} 
for the years 2002-2006.

%%%%%%%%%%%%%%%%%%%%%%%%%%%%%%%%%%%%%%%%%%%%%%%%%%%%%%%%%%%%%%%%%%%%

%%%%%%%%%%%%%%%%%%%%%%%%%%%%%%%%%%%%%%%%%%%%%%%%%%%%%%%%%%%%%%%%%%%%

Afterward, we extracted the 03:00 UTC (BLH5), 9:00 (BLH7) 
and 21:00 UTC (BLH11) data from the retrieved BLH fields. 
The CWT classification was determined for Portugal according to Trigo and DaCamara \cite{trigocamara} as described in Sec.~\ref{sec:cwt}. Values for daily mean sea level pressure (SLP), relative humidity and temperature and geopotential height at the 1000 hPa level values were extracted from ERA Interim Reanalyses dataset \citep{Dee} on a grid of $1^o$ latitude by $1^o$ longitude for Portugal (40W-30E, 20-70N). The period between 1981 and 2010 was used to perform a 30 year climatology that included the air quality period under analysis (2002-2006). Based on the large-scale fields, prevailing CWTs at regional scale were determined using the simple Geostrophic approximation according to the methodology proposed by Trigo and DaCamara \cite{trigocamara}. The daily CWTs resulting from the classification procedure were then considered as an input variable.

In total there are $15$ variables that are available as input data for the NN model.  
Table \ref{fig_tabela_var} summarizes the input training data.

Based on the available five years datasets, we constructed a collection of 
records, consisting of the input vector, which included the meteorological 
variables, air pollutant concentrations, and the corresponding target \pa.
All the data used refers to the period between 1/1/2002 and
31/12/2006. The first four years were used to construct the 
models and the year 2006 was used for independent evaluation
(see Sec.~\ref{sec:forecast}).

%%%%%%%%%%%%%%%%%%%%%%%%%%%%%%%%%%%%%%%%%%%%%%%%%%%%%%%%%%%%%%%%%%%%%%%%%%%%%%%%
\section{Methods}
\label{sec:methods}

\subsection{Circulation-to-environmental approach}
\label{sec:cwt}

The concentration of pollutants in the atmosphere are linked to the occurrence 
of certain synoptic weather conditions \citep{demuzere} and to the regional wind 
flow pattern induced by mesoscale meteorological processes (land-sea breezes) 
\citep{DayanLevy2004}. CWT dictates the long-range transport, linking a particular air mass 
to dispersion conditions and also to the mesoscale meteorological configuration 
that controls the regional transport of air pollution \citep{DayanLevy2004}.
Considering the capabilities of this approach, these prevailing circulation patterns have witnessed a growing interest by the research community during the last decades \citep{demuzere, trigocamara}. The aim of these studies varies considerably, ranging from applications to climatic variability, including trends and extreme years, to environmental purposesand also to access weather driven natural hazards.

CWTs objective classification has successfully been applied to Portugal mainland 
by Trigo and DaCamara \citep{trigocamara}, who linked CWTs to precipitation. 
Pereira et al. \citep{Pereira2005} and Ramos et al. \citep{Ramos2011} analysed the 
impacts of atmospheric circulation, respectively, on fire activity and on lightning 
activity over Portugal. 
There are other studies within the Iberian Peninsula, most of them 
focusing on climatic trends \citep{Lorenzo2008}, associated to extreme events or 
to extreme years \citep{garciaherrera2007, VicenteSerrano2011}.

The majority of CWT classification procedures are based on the application of statistical 
selection rules (e.g.cluster analysis, regression trees), but can also be
based on the determination of physical parameters regarding the prevailing 
atmospheric circulation pattern. Furthermore, CWTs are generally specific to a given region, resulting from the examination 
of synoptic weather data (e.g.~sea level pressure (SLP) or geopotential height 
at 500 hPa) \citep{Ramos2011}.

In this paper, prevailing CWTs calculated according to Trigo and DaCamara 
\citep{trigocamara} are considered as a potencial predictor.

\subsection{Predictors choice}

A crucial step in the development of a forecast model is the choice of input 
parameters, the predictors \citep{Hooyberghs2005}. 
Predictors can be fed into a model for training and evaluation in numerous ways. 
Usually, a number of statistical methods can be applied 
in order to choose the most appropriate set of predictors/ inputs.
Important methods in this scope are stepwise regression (SR), principal component analysis 
(PCA), cluster analysis and ARIMA \citep{wilks2006}. 
These methods are pre-processing procedures, which allow reducing the number of 
input variables into the models, thus eliminating redundant information, instabilities and over-fitting.

Here, the selection of variables was made independently for each monitoring station 
through a forward stepwise regression (FSR), from which the best time lag for each 
input variable was also determined.
%The BSR allows variables to be optimized in order to predict \pa\ at each monitoring 
%station. 
During this procedure, which starts with the variable most correlated with the target, new variables are added which, together
 with the old one(s), most accurately predicts the target\citep{wilks2006}.
The procedure stops when any new variable does not 
significantly reduce the prediction error. Significance is measured by a 
partial F-test applied at 5\% \citep{Wong,wilks2006}.

\subsection{The Neural Network framework}

Neural networks (NN) are mathematical models inspired by the biological nervous 
system \cite{Cobourn2000,Agirre-Basurko2006,Gardner1998}, since they are composed 
by a number of interconnected entities, the artificial neurons, which are similar in several ways to biological neurons. 

\begin{figure}[t]
  \centering
  \includegraphics[width=0.5\textwidth]{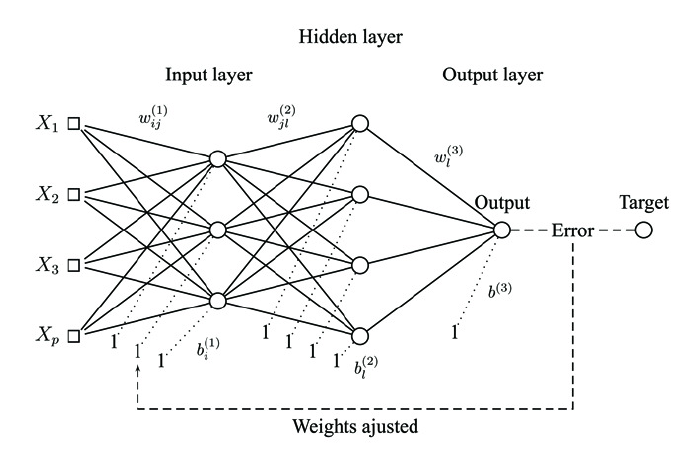}
  \caption{\protect 
           Illustration of a feed-forward artificial 
           neural network model with three layers. 
           Input variables $X_{i=1,...,p}$ can be perceived as
           ``neurons'', $\omega$ represent the weights associated 
           to each neuron and $b$ are the bias vectors which combined 
           will produce an output within certain error limits. 
           (Adopted from Russo et al. \citep{russo}).}
\label{fig3}
\end{figure}

One of the most common examples of architectures used is the multilayer 
perceptron \cite{Agirre-Basurko2006, Haykin1999}, where the artificial neurons can be organized following different types of architectures, 
composing a certain number of levels (Fig.~\ref{fig3})\citep{Agirre-Basurko2006, Haykin1999}. 
In the zero level one has the set of independent variables, $X_i$, and a number of
connections with a weight $\omega_{ij}$, joining the variables $X_i$ to neurons in 
the next level \citep{Gardner1998,TrigoPalutikof1999}. In the first level (``input layer'' in 
Fig.~\ref{fig3}), each neuron computes a linear combination of the weighted 
inputs $\omega_{ij}$, including a bias term $b_j$: $Y_j = \sum_i \omega_{ij}X_i + 
b_j$. This sum is transformed using a linear or non-linear function, $W_j=f(Y_j)$. 
The weights can be initially randomly chosen, and are then properly tuned during
the training of the NN as described below.
The bias term is included in order to allow the activation functions to be offset 
from zero and it can be set randomly or set to a desired value, such as a dummy 
input with a magnitude equal to one. 

The output $W_j$ obtained at 
the previous level is then passed as an input to other 
nodes in the following layer, usually named hidden layer. This procedure is 
performed repeatedly to better tune the weights until a certain accuracy threshold 
between the produced output and the target variable (empirical data) is reached.
It is possible to use several hidden levels, sucessively. However, it is often 
advantageous to minimize the number of hidden nodes and layers, in order to improve 
the generalization capabilities of the model and also to avoid 
over-fitting \cite{Gardner1998}.
%%%%%%%%%%%%%%%%%%%%%%%%%%%%%%%%%%%%%%%%%%%%%%%%%%%%%%%%%%%%%%%%%%%%
%%%%%%%%%%%%%%%%%%%%%%%%%%%%%%%%%%%%%%%%%%%%%%%%%%%%%%%%%%%%%%%%%%%%

There are several training procedures for estimating the weights and 
associate input and output. 
Here, we use a modified version of the back-propagation (BP) procedure, 
which is one of the most popular and common training procedures, 
 see e.g.~\citep{Haykin1999,TrigoPalutikof1999}.
As any other training algorithm, BP has drawbacks.
One is that the convergence may be slow and the final weights may be trapped 
in local minima over the highly complex error surface \citep{Haykin1999,TrigoPalutikof1999}. 
To overcome this shortcoming, numerically optimized techniques have been developed, 
such as the Levenberg-Marquardt method (LM) which are based on an approximation 
of the Gauss-Newton method.
The LM method has the advantage of converging faster and with a higher
robustness than most back-propagation (BP) schemes \citep{TrigoPalutikof1999} 
because it avoids computing second-order derivatives.

However, the Levenberg-Marquardt requires more memory \citep{Haykin1999, 
TrigoPalutikof1999} than other simpler training procedures, as for instance, the 
Widrow-Hoff learning-rule, also known as 
the least mean square (LMS) rule, which is usually used for training linear models.
Unlike the standard BP algorithm that can be trapped in local minima, the 
Widrow-Hoff rule will give a unique solution corresponding to the absolute minimum 
value of the error surface \citep{Haykin1999}.
For this reason, we chose to use the Widrow-Hoff rule for the linear 
approach.

As stated above we will consider linear and non-linear NN.
The linear NN model is composed of a single one-layer NN structure with just 
one neuron, which employs a linear activation function and behaves exactly 
like a linear model, producing the same results as a linear regression 
model \citep{Weisberg1985}. Following other previous approaches, this type 
of NN structures constitute the baseline against which the performance of 
the non-linear model will then be compared. 
The non-linear NN models used here are based on a feed-forward configuration of the multilayer perceptron that has been used by several authors \citep{Hooyberghs2005, Papanastasiou}. %We have tested a large number of architectures; however the use of 2 layers was verified to be sufficient. The use of a 3rd layer was shown to be redundant. 
Additionally, we used a linear transfer function in the only node of the output layer, and the log-sigmoid function in the nodes of the other layers.

For the sake of simplicity we will refer from this point forward to the 
linear model as MLR and to the non-linear model as NN solely.

\subsection{Application of the NN framework}

The NN framework is applied to our data set in the following way.
Consider an attribute $Z(x,t)$, symbolizing the concentration of
\pa, measured at a spatial location $x$ at day $t$, 
which yields a daily series of the pollutant's concentration 
at each monitoring station.

One considers then its decomposition into a periodic component $M(x,t)$ and a 
residual $R(x,t)$, yielding $Z(x,t)=M(x,t)+R(x,t)$. 

In particular we consider the periodic components $M(x,7)$, which is determined respectively by a 7 days 
moving average.
Likewise, we take also the respective residuals $R(x,7)$ obtained from the removal of the correspondent periodic 
component.

We then apply the linear and non-linear models, i.e.~MLR and NN
models, to each monitoring station $x$ in order to model
both the complete signal, hereafter called \textbf{TOT} approach, and
to model the residual components, hereafter called \textbf{RES}
approach.
The forecasting capabilities of the different approaches are 
compared in order to assess the potential improvement using non-linear 
NN in air quality modeling.

The NN models used here are based on a feed-forward configuration of 
the so-called multilayer perceptron \cite{russo}, sketched in 
Fig.~\ref{fig3}, that has been used by several authors 
\citep{Hooyberghs2005, Papanastasiou,russo, TrigoPalutikof1999}. 
We tested large number of architectures, each one with a given number
of layers. The use of two layers was verified to be sufficient, 
since a superior number of layers does not improve the output.

For the linear model MLR, a perceptron with a linear activation function 
was used, while for the non-linear NN models, 
the log-sigmoid function was used, except for the single node in
the output layer, for which we consider a linear transfer function.

The MLR models are trained with the LMS rule and the NN models with 
the Levenberg-Marquardt method. In both cases, a cross-validation is
applied with the available period being divided into four times and the 
calibration-validation procedure is completed four times independently, 
i.e.~each time three years are used for construction and the remaining 
year is retained for validation. 
Further, a moving window is applied.
Thus, the first run is performed 
using data for 2002-2004 to train the model, whereas data from 2005 is 
used for validation purposes. In the second run, data for 2003-2005 are 
used for training and data for 2002 for validating, and so on.

With such cross-validation procedures \citep{wilks2006}, it is possible to account for the risk of over- or underfitting.
Moreover, in this way, 
one is able to ascertain if the models are stable and if they 
are capable of generalizing correctly in forecast mode.
After models’ calibration and validation with 
historical data (2002-2005), 
the models are used to produce forecasts for the daily average of \pa\ 
concentration, during a period of one year.
For this purpose an independent one-year sample, the year 2006,
is left out in order to be used for evaluation of models performance 
(see Sec.~\ref{sec:performance}) during the individual daily average 
predictions. 

In the end, the forecasts are then compared with the actual observed 
pollutant values at the monitoring stations.

\subsection{Performance indicators}
\label{sec:performance}

Rigorous quantitative measures are required to perform models’ evaluation. 
Thus, in order to evaluate the efficiency and performance of the developed 
models three continuos performance indicators are used. 
The simplest measure is the the Pearson correlation coefficient (PC): 
\begin{equation}
\hbox{PC} = \frac{ \sum_{i=1}^N(y_i-\bar{y})(o_i-\bar{o}) }
           {\left[ \sum_{i=1}^N(y_{i}-\bar{y})^2\sum_{i=1}^N(o_{i}-\bar{o})^2 \right]^{1/2}} ,
\label{PC}
\end{equation}
where $y_i$ denotes the respective model forecast at time $i$
 while $o_i$ denotes the real 
observed values at time $i$, and $\bar{y}$ and $\bar{o}$ are the corresponding average
values.

A quantity similar to PC, also related to correlation between series, is the root mean square 
error (RMSE) given by
\begin{equation}
\hbox{RMSE} = \sqrt{\sum_{i=1}^N(y_i-o_i)^2}.
\label{RMSE}
\end{equation}

Considering that correlation coefficients are not robust to deviations from linearity, 
its exclusive use to evaluate the quality of a model can lead to misleading 
results \citep{wilks2006}. Therefore we consider these quantities combined with other 
properties which present different abilities for
accessing important aspects of the data such as outliers and average values.

The skill against persistence, SSp, which is interpreted as the 
percentage of improvement that our model can provide when compared with the 
persistence model \citep{wilks2006,TrigoPalutikof1999}, i.e.~the model that yields the 
observed value of yesterday as the forecast for today. The score is quantitatively defined as
\begin{equation}
\hbox{SSp} = \frac{\frac{1}{N}\sum_{i=1}^N(y_i-o_i)^2-\frac{1}{N-1}
\sum_{i=1}^N(o_{i+1}-o_i)^2}{\frac{1}{N-1}\sum_{i=1}^N(o_{i+1}-o_i)^2}\, .
\label{skillpers}
\end{equation} 
The SSp is also used as a measure of the relative accuracy of the model.

Both linear and non-linear models will be compared with this persistence 
model, which is the simplest way of producing a forecast and assumes that 
the conditions at the time of the forecast will not change. Due to a certain 
level of memory that characterizes air pollutants, persistence corresponds 
to a benchmark model considerably more difficult to beat than climatology or 
randomness \cite{demuzere}.

%The second skill score is called the coefficient of efficiency, CE, and
%evaluates the model capability to predict values that are quite distant 
%from the mean value. This coefficient is sensitive to large differences 
%between observed and modelled values and is overly sensitive to extreme
%values \citep{legates}. 
%Quantitatively the implementation of CE reads:
%\begin{equation}
%\hbox{CE} = \frac{\sum_{i=1}^N(y_i-o_i)^2}{\sum_{i=1}^N(o_{i}-\bar{o})^2}.
%\label{CE}
%\end{equation} 
%Large values of CE therefore mean that the model is robust and that is able to produce forecasts that will be meaningful in forecast mode.

Additionally, four categorical measures are also considered, 
to ascertain if the models are able to predict 
exceedances \citep{wilks2006}. 
Traditional categorical metrics used in model evaluations assess the 
model’s ability to predict an exceedance which is defined by a fixed 
threshold. 
These metrics are defined by sets of observational forecasts %sets 
that are paired togheter. Here, we used the false alarm rate (F), 
i.e.~the proportion of non-occurrences incorrectly forecasted, and the 
proportion of correctness (PCS), i.e.~the proportion of events properly 
forecasted.
Both categorical measures, F and PCS, are applied against binary time 
series obtained with thresholds for poor air quality limit values 
(\pa-50 $\mu g/m^3$),
defined by the Portuguese National Environmental Agency.
One should however notice that
there is now considerable evidence that daily hospital admissions for 
cardiorespiratory diseases are linked to levels of \pa\ not only on the same, but also on  
previous days \citep{Wong} and that association is positive for values 
lower than the legal thresholds.
Thus, two additional categorical measures were introduced in order to 
assess if the models are able to performe correctly for a new threshold 
that corresponds to 50\% of the legal limit value (F50 and PCS50).
%%%%%%%%%%%%%%%%%%%%%%%%%%%%%%%%%%%%%%%%%%%%%%%%%%%%%%%%%%%%%%%%%%%%%%%%%%%%%%%%%%%%%
\begin{table*}[t]
\center
\footnotesize
%\begin{tabular}{|c|c|c|c|c|c|c|c|c|c|c|c|c|c}
\begin{tabular}{cccccccccccccc}
\hline
&\multicolumn{13}{c} {\textbf{Stations}} \\[3pt]
    \cline{3-14}\\& & E &O &AL & L & ESC& R & LAR & LRS & CC& QM & MM &OD\\[3pt]
\hline
\multirow{2}{*}{\pa}&\textbf{TOT} & 5&2& 2& 2& 3& 2& 2& 6& 2& 3&4&7\\
										\cline{2-14}
										& \textbf{RES}-7 & & & & & & &4& & 9& & & \\
										\hline
\multirow{2}{*}{\pa 0h UTC}&\textbf{TOT} & 1&1&1&1&1&1&1&1&1&1&1& 1\\
										\cline{2-14}
										& \textbf{RES}-7 & 1&1&1&1&1&1&1&1&1&1&1& 1\\
										\hline
\multirow{2}{*}{CO}&\textbf{TOT} & & & & & 5&9& 7&9& && 5& \\
										\cline{2-14}
										& \textbf{RES}-7 &5 & & && 4&3& &9&7&7& 6&\\
										\hline
\multirow{2}{*}{\no}&\textbf{TOT} & & & 7& 8&6& & & 10& & & & 3\\
										\cline{2-14}
										& \textbf{RES}-7 & 3&4&2&& 2& & & 8& 8& 6&8&3\\
										\hline
\multirow{2}{*}{NO}&\textbf{TOT} &  & &8 & & & & & & 4& 5& &4\\
										\cline{2-14}
										& \textbf{RES}-7 &  & & & & & & & & & 5& &5\\
										\hline
\multirow{2}{*}{\pa m}&\textbf{TOT} & 6&7& & & & 8& & 7& & 7&7 &6\\
										\cline{2-14}
										& \textbf{RES}-7 & & & & & & & & 7& & & & 4\\
										\hline
\multirow{2}{*}{$V_d$}&\textbf{TOT} & 3&4&3&4& & 4& 4&3&3& 4& 3& 8\\
										\cline{2-14}
										& \textbf{RES}-7 & 4&3&3&3& & 4&3& 3& 2& 2&3&\\
										\hline
\multirow{2}{*}{$V_i$}&\textbf{TOT} & & & & &7 & & & & & & & \\
										\cline{2-14}
										& \textbf{RES}-7 & & & & & & & & & & & & \\
										\hline
\multirow{2}{*}{Rad}&\textbf{TOT} & & & &5 & & & & & &9 & & \\
										\cline{2-14}
										& \textbf{RES}-7 & & & & & & & & & & & & \\
										\hline
\multirow{2}{*}{Hum}&\textbf{TOT} & 2&3&4&7&2&3&3&5& &  & & \\
										\cline{2-14}
										& \textbf{RES}-7 & 2&2& &2& & 2&2&2&6& & 2& 2\\
										\hline
\multirow{2}{*}{Tmax}&\textbf{TOT} & 4&5&6&3& & 6&5& 2& 5&2&2&2\\
										\cline{2-14}
										& \textbf{RES}-7 &  & & & 6& &  & 6& 5& 3& 3& 5&7\\
										\hline
\multirow{2}{*}{CWT}&\textbf{TOT} &  & & & & & & & & & & &  \\
										\cline{2-14}
										& \textbf{RES}-7 & 7 & & &5& & & 7& & 5&9& 9&8\\
										\hline
\multirow{2}{*}{BLH5}&\textbf{TOT} & 7&8& 5 &  & 7 & & 8& 6& 6&  & \\
										\cline{2-14}
										& \textbf{RES}-7 & & & & & & & & & 4& 4&  &\\
										\hline
\multirow{2}{*}{BLH7}&\textbf{TOT} &  & & &  &4 & & & & & & &5\\
										\cline{2-14}
										& \textbf{RES}-7 & &5& & & 3&5& & 4& & & 4 &6\\
										\hline
\multirow{2}{*}{BLH11}&\textbf{TOT} & 8&6& & 6& & 5& 6& 4& & 8& 6& \\
										\cline{2-14}
										& \textbf{RES}-7 & 6& & & 4&  & & 5& 6& & 8& 7&\\
										\hline
\end{tabular}
\caption{\protect 
           Available and chosen predictors by FSR for each monitoring station.}
\label{fig_tabela_resultados}
\end{table*}
%%%%%%%%%%%%%%%%%%%%%%%%%%%%%%%%%%%%%%%%%%%%%%%%%%%%%%%%%%%%%%%%%%%%%%%%%%%%%%%%%%%%%

%%%%%%%%%%%%%%%%%%%%%%%%%%%
\section{Results and Discussion}
\label{sec:results}

\subsection{Selection of input variables}

We first consider all 15 potential predictors %were considered for each pollutant 
for PM$_{10}$ (see Table~\ref{fig_tabela_var}). 
The use of the FSR has reduced the complexity by retaining substantially less 
variables, namely only those marked in Table~\ref{fig_tabela_resultados}.

We also found that adding time lags superior to one day do not provide relevant 
additional information. Therefore, only the one-day time lags for both meteorological 
and air quality variables are taken into account in the subsequent analysis. 

Our analysis further revealed that the most significant variable in predicting \pa\ 
for all the monitoring stations is the 0h UTC \pa\ concentration. 

Other variables that were retained for the majority of the sations under the \textbf{TOT} approach are the previous day average and 
maximum \pa\ concentrations, the previous day average values of \no\, NO and CO concentrations, 
the maximum temperature, wind direction, humidity and BLH. 
The other variables retained for the majority of the sations under the \textbf{RES}-7 approach are the previous day average values of \no\ and CO concentrations, the maximum temperature, wind direction, humidity, CWT and BLH.
Additionally, the \textbf{RES}-7 approach uses considerably less pollutant related predictors than the \textbf{TOT} approach, including also the CWT classification as one of the most important predictors in the majority of the monitoring stations. 

While the dependence on the wind, 
relative humidity and BLH were also shown in previous works \citep{Hooyberghs2005, demuzere},
the \no\ and CO dependence is present due to road traffic influence, as road traffic 
behaves as a local source of PM$_{10}$ \citep{demuzere}.

Kukkonen and co-workers \citep{kukkonen} showed that the inclusion of meteorological 
variables for the day of prognosis improves the performance of NN models and that linear 
models perform significantly worse in this situation. However, we consider that these 
variables might unnecessarily increase the error associated to the prediction and choose 
not to include them at this stage.

%%%%%%%%%%%%%%%%%%%%%%%%%%%%%%%%%%%%%%%%%%%%%%%%%%%%%%%%%%%%%%%%%%%%%%%%%%%%%%%%%%%%%%%%%%%%%%%
\begin{table*}[ht]
\begin{center}
		\begin{tabular}{l | l | c c c c c c c }
& \small 	Model& \small PC& \small 	SSp&  \small RMSE& \small 	F	& \small PCS& \small 	F50& \small 	PCS50\\
\hline
\small PM$_{10}$	& \small {\textbf{TOT}}-MLR& \small 	0.75& \small 	45.00&  \small 12.85	&\small 6	& \small 88	& \small 50& \small 	80\\
 & \panel{l}{\small 	{\textbf{RES}-7}-MLR}	& \panel{c}{\small 0.81}	& \panel{c}{\small 54.41}&  \panel{c}{\small 11.69}	&\panel{c}{\small 12}	& \panel{c}{\small 86}	& \panel{c}{\small 62}	& \panel{c}{\small 89}\\
	& \small {\textbf{RES}-7}-NN2	& \small 0.81	& \small 54.30& \small 	11.69& \small 	11& \small 	85& \small 	64& \small 	90\\
& \small {\textbf{RES}-7}-NN3	& \small 0.81	& \small 54.20&  \small 	11.69& \small 	11	& \small 85& \small 	63	& \small 90\\
% & \small 	{\textbf{RES}-30}-MLR	& \small 0.76	& \small 48.21& \small 	0.56	& \small 12.46	&\small 7	& \small 89	& \small 47	& \small 81\\
%	& \small {\textbf{RES}-30}-NN2	& \small 0.73	& \small 44.56& \small 	0.53&  \small 	12.80&  \small 	6&\small 	89& \small 	48& \small 	80\\
%	& \small {\textbf{RES}-30}-NN3	& \small 0.73	& \small 43.92& \small 	0.53&  \small 	12.83& \small 	7	& \small 89& \small 	48	& \small 80\\
\hline%\\%[4pt]%\hline			
\end{tabular}
	\end{center}
\caption{\protect 
           Average performance indicators obtained for the \pa\ calibration/validation process, including the Pearson correlation coefficient (PC), the skill against persistence (SSp (\%)), the coefficient of efficiency (CE), the root mean square error (RMSE ($\mu gm^{-3}$)), the false alarm rate (F (\%)), the proportion of correctness (PCS (\%)) ,  the 50\% false alarm rate (F50 (\%)),  and the 50\% proportion of correctness (PCS50 (\%)). Each average performance indicator was determinded based on the indicators of all the monitoring stations.}
\label{fig_tabela_resultados2}
\end{table*}
%%%%%%%%%%%%%%%%%%%%%%%%%%%%%%%%%%%%%%%%%%%%%%%%%%%%%%%%%%%%%%%%%%%%%%%%%%%%%%%%%%%%%%%%%%%%%%%%%%%%%

\subsection{Comparison of methods}

The validation tests presented here are based on the use of MLR and NN models in which all the 
retained predictor variables are incorporated according to Table 2 framework. Validation results obtained with the MLR and 
NN models are shown in Table 3. The numbers of hidden neurons applied are identified by the 
numeric index after NN, i.e., NN2 refers to a NN model with 2 neurons in the hidden layer. 
The choice of the number of hidden units was made iteratively. There are four main conclusions to be drawn from Table \ref{fig_tabela_resultados2}:

\begin{enumerate}
	\item All the models perform substantially better than persistence with SSp scores above 45\%.

	\item The proportion of correctness (PCS and PCS50) 
are quite high which indicate that the models are robust and are able to correctly predict not only
medium values but also events with high values.

	\item The false alarm rate (F) is significantly low for high values, which indicates that the only a low percentage of pollution episodes are not correctly predicted relatively to the legal limit.

	\item Weekly residuals ({\textbf{RES}}-7) models outperform the \textbf{TOT} models. Removing the weekly cycle appears to be a promising approach compared to the complete signal model (\textbf{TOT}).

\item The {\textbf{RES}}-MLR model performs approximately the same than {\textbf{RES}}-NN2 and {\textbf{RES}}-NN3, 
and considerably better than \textbf{TOT}-MLR models. The similitude between {\textbf{RES}}-NN and {\textbf{RES}}-MLR 
results that it looks doubtful that there is a significant advantage in using the NN 
model comparatively to the MLR in the present case. 
\end{enumerate}

These findings altogether indicate that there is no significant advantage in the use of NN 
against MLR. Henceforth, we will restrict the remaining analysis to the MLR approach.

%\textbf{RES}-7-MLR shows also the largest effectiveness CE.
From the operational point of view, the effectiveness of a prediction model should be judged 
according to its ability to forecast properly in order to be able to alert the population and 
the competent health authorities. However, the forecast models are known a priori to be imperfect, 
thus the alert threshold must be set below the critical level objectively identified, in order 
to allow for a margin of safety \citep{Cobourn2000}.

Still, the performance indicators presented here are superior to those obtained by Demuzere 
and co-workers \cite{demuzere} for the Netherlands, and are consistent with the results presented 
by Hooyberghs and colleagues \cite{Hooyberghs2005} for Belgium. Demuzere et al. \cite{demuzere} presented a 
performance of 
%$PC=\sqrt{0.42}$
$PC=0.648$
for particulates and SSp 
lower values achieved through {\textbf{RES}}-MLR. Hooyberghs et al. \cite{Hooyberghs2005} presented results of $PC$ between 
$0.65$ and $0.80$ for \pa.
Here we attain similar results, $0.75<PC<0.81$, by incorporating meteorological variables. Moreover, checking the performance results, one observes a tendency for higher performance 
for the independent validation, which is due to the favourable characteristics
of year 2006, as we explain in the next section.
%%%%%%%%%%%%%%%%%%%%%%%%%%%%%%%%%%%%%%%%%%%%%%%%%%%%%%%%%%%%%%%%%%%%
\begin{figure}[t]
  \centering
  \includegraphics[width=0.5\textwidth]{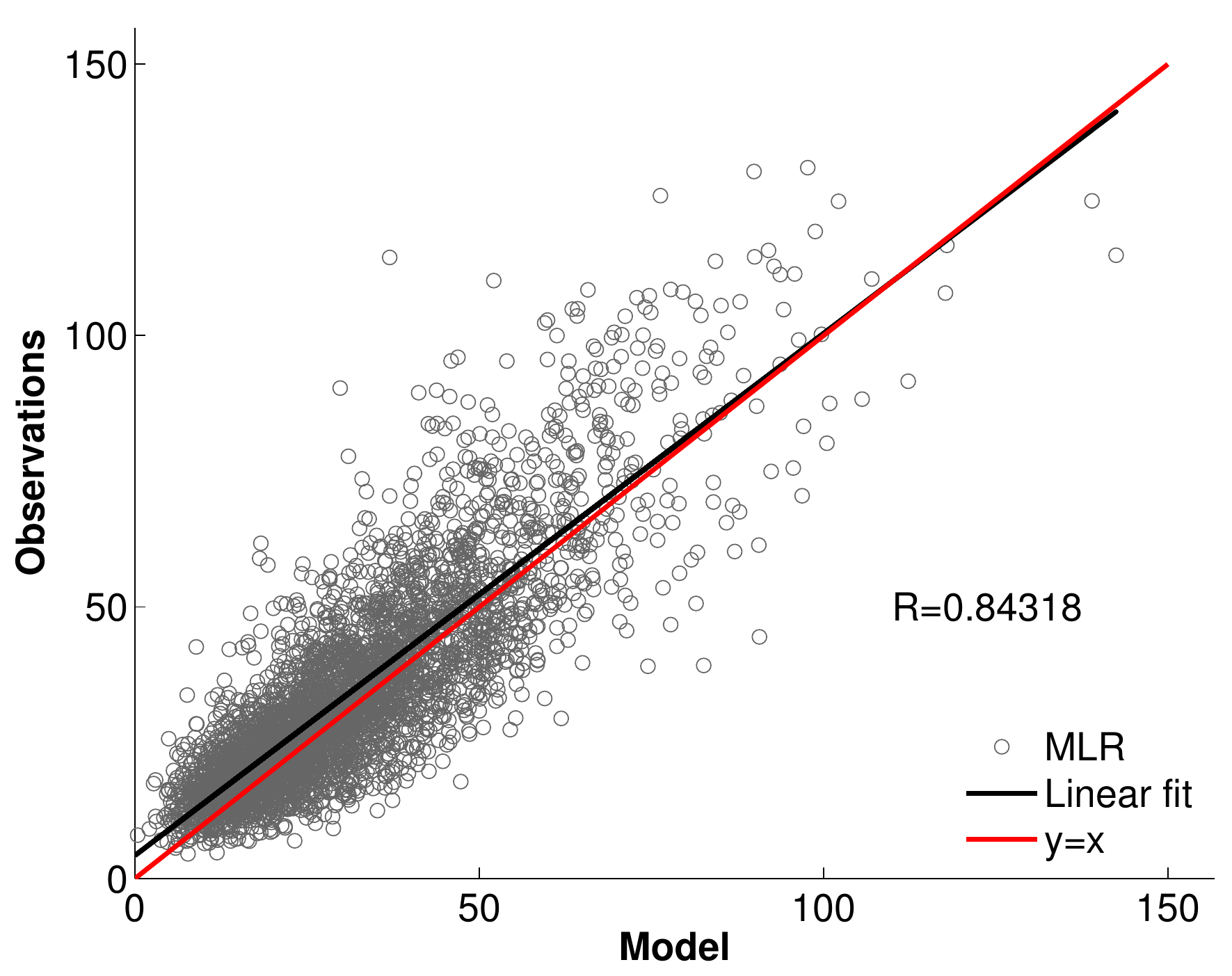}
  \caption{\protect Scatter plots of MLR results versus actual observed 
           \pa\ values for all monitoring stations and for the year 2006.}
\label{fig4}
\end{figure}
%%%%%%%%%%%%%%%%%%%%%%%%%%%%%%%%%%%%%%%%%%%%%%%%%%%%%%%%%%%%%%%%%%%%
%%%%%%%%%%%%%%%%%%%%%%%%%%%%%%%%%%%%%%%%%%%%%%%%%%%%%%%%%%%%%%%%%%%%
\begin{table}[htb]
\centering
\begin{tabular}{l | c  c | c}
%			\includegraphics[trim=150 300 150 10,clip, scale=1.1]{figs/forecast_7.pdf}%}
%\small \textbf{Station }&\multicolumn{2}{| c |}{\small PM$_{10}$} \\
\small {Station }& 2002-2005 & 2006 & $\% \Delta$\\
\hline
\small {E}	& \small 	0.83	& \small\bfseries (\textbf{0.78})& -5\\
\small {O}	& \small 	0.79	& \small\bfseries (\textbf{0.86})&7\\
\small {AL}&  \small 0.81& \small\bfseries 	(\textbf{0.82})&1\\
\small {L}	& \small 0.83	& \small\bfseries (\textbf{0.86})&3\\
\small {ESC}		& \small 0.80	& \small\bfseries (\textbf{0.83})&3\\
\small {R}	& \small 	0.79& \small\bfseries 	(\textbf{0.87})&8\\
\small {LAR}	& \small 	0.85	& \small\bfseries (\textbf{0.87})&2\\
\small {LRS}	&\small 	0.83	& \small\bfseries (\textbf{0.87})&4\\
\small {CC}	& \small 	0.75& \small\bfseries	(\textbf{0.78})&3\\
\small {QM}	& \small 	0.83& \small\bfseries 	(\textbf{0.86})&3\\
\small {MM}	& \small 	0.82& \small\bfseries 	(\textbf{0.86})&4\\
\small {OD}	& \small 	0.85& \small\bfseries 	(\textbf{0.82})&-3\\
%-	Model wasn't developed due to the lack of information 
%\hline
\end{tabular}
\caption{\small\normalfont 
         Correlation coefficients between observed and modelled \pa\ concentrations for each station considered and for the calibration/validation period (2002-2005) and for the independent forecast year (2006).} 
\label{tab:corr} 
\end{table}
%%%%%%%%%%%%%%%%%%%%%%%%%%%%%%%%%%%%%%%%%%%%%%%%%%%%%%%%%%%%%%%%%%%%%%%%%%%%%%%%%%%%%%%%%%%

%%%%%%%%%%%%
\subsection{Forecast: Independent validation}
\label{sec:forecast}

The forecasts retrieved by the MLR models were compared with the actual observed pollutants 
values of the year 2006 at the monitoring stations.
The scatter plots and correlation coefficients between observed and modelled values were 
computed for all monitoring stations.
Figure \ref{fig4} presents the aggregated scatter plots and correlation coefficient for all 
monitoring stations. 
The results for the independent sample show a very high average correlation ($PC> 0.84$) between 
the predicted and observed values.

In Table~\ref{tab:corr} the correlation coefficients for each individual monitoring station
for the calibration/validation period (2002-2005) and for the one-year independent sample 
(2006) are presented. These results show that MLR model generalizes well for independent data 
and for each monitoring station. 

In general, MLR techniques are known to underestimate peak levels. Interestingly, although the 
MLR model is built using the calibration dataset only, we can observe an increase in accuracy 
for the majority of the stations when in forecast mode. This may be explained by the 
characteristics of the historical data used to construct the models: The year 2005, 
which was included in the construction of the model, is considered an atypical meteorological 
year, with low wind and high temperatures and with a prolonged drought \citep{garciaherrera2007}.
On the other hand, the \pa\ data sets used on this work comprehend the years from 2002 to 
2006 in Lisbon. For this location, the years of 2003 and 2005 were particularly outstanding 
relatively to weather conditions, namely an exceptional heatwave that struck the entire western 
Europe, in 2003 \citep{trigo2006} and one of the most severe droughts of the 20th century occurred in 2005 \citep{garciaherrera2007}.

Moreover, air pollution is strongly influenced by shifts in the weather. Changes in the 
temperature, humidity and wind indeed induce changes in the transport, dispersion, and
transformation of air pollutants at multiple scales \citep{Dias}.
Therefore, using all the years as individual calibration/validation
samples, yields quite disparate skill values on one hand with an
average that is significantly below the skill against persistence
obtained when using these anomalous years for independent
validation of 2006. 

%%%%%%%%%%%%%%%%%%%%%%%%%%%%%%%%%%%%%%%%%%%%%%%%%%%%%%%%%%%%%%%%%%%%%%%%%%%%%%%%%%%%%%%%%%%%%%%%%%%%
\section{Conclusions}
\label{sec:conclusions}

In this paper we introduce a framework consisting in a pre-selection procedure of predictors
which are then used as input data to train NN model.

In order to assess the importance of the periodic and residual components present in 
pollutants time series, the application of linear (MLR) and non-linear (NN) models to 
each monitoring station was performed. 
The forecasting capabilities of the different approaches were then compared.
The approach based on the removal of the weekly cycle presented the best results, 
comparatively to the use of the complete signal. 
Moreover, MLR and NN showed similar performances when evaluated by 
each of the above criteria. Therefore we find it reasonable to conclude that there is no 
significant advantage on the use of NN against MLR for the case studied.
 
Linear MLR and non-linear NN models designed to forecast daily average \pa\ concentrations 
in Lisbon, Portugal, were used to produce forecasts and hindcasts. The models were calibrated 
using air quality and meteorological data from 2002 until 2006 taken at 12 pollutant 
monitoring stations.

Our framework enables to rank all given variables, and then select the highly ranked variables
as predictors, which were chosen for each monitoring station separately. To rank the
variables a forward stepwise regression was used.
We found that the most significant variables in 
predicting \pa\ are pollutants related to road traffic emissions and meteorological 
variables related to atmospheric stability. Particularly for the \textbf{RES}-7 approach, 
the most significant variables in predicting \pa\ are, in descending order of importance, the 
0h UTC \pa\ concentration, the previous day average values of \no\ and CO concentrations, the 
maximum temperature, wind direction, humidity, CWT and BLH. These results enphasize the 
importance of meteorological variables and of the circulation-to-environment approach to air 
quality forecast.

In particular, we found that for forecasting \pa in Lisbon, CTW should be taken 
as input data, though its rank is not particularly high compared with other meteorological
data. 
However, we point out that the ranking of predictors varies considerably from one station
to another, since it reflects the diversity of geographical and urban features, such
as traffic, industries, distance to the coast.
Therefore, a forthcoming approach to urban pollution would be to apply such procedure
to a panoply of different pollutants and ascertain which ones are more sensible to 
synoptic scale circulation and meteorological constraints.
Another issue to addressed in a forthcoming study is the interaction between stations.

All in all,
the models presented here and the introduced framework
are able to produce different results for each 
monitoring station, which allows a good spatial resolution for Lisbon’s 
urban area. Consistent with the performance measures, high pollutants’ 
peak values were reproduced in most cases by each model. The simplicity 
and cost efficiency of these models, associated with their performance 
capabilities, show to be very promising for urban air quality characterization, 
allowing further developments in order to produce an integrated air quality 
surveillance system for the area of Lisbon. 
Being a general numerical procedure for any given set of measurements,
our finding can be easily adapted to other NN models in weather or
geophysical forecast. An extension of this work to take into account 
the correlations between a higher number of measurement stations is planned.
%
%%%%%%%%%%%%%%%
\section*{Acknowledgments}

The authors acknowledge the {\it Instituto de Meteorologia} and the
{\it Ag\^encia Portuguesa do Ambiente} for the meteorological and environmental 
data, respectively.
The authors thank DAAD and FCT for financial support through 
the bilateral cooperation 
DRI/DAAD/1208/2013.
FR (SFRH/BPD/65427/2009) 
thanks {\it Funda\c{c}\~ao para a Ci\^encia e a Tecnologia}
for financial support, also with the support 
Ref.~PEst-OE/FIS/UI0618/2011.
PL thanks BMU (German Environment Ministery) under the project 41V6451.

%%%%%%%%%%%%%%%%%%%%%%%%%%%%%%%
%%% BIBLIOGRAPHY %%%%%%%%%%%%%%
%\end{linenumbers}

\bibliographystyle{elsarticle-num}

%%%%%%%%%%%%%%%%%%%%%%%%%END BIBLIOGRAPH%%%%%%%%%%%%%%%%%%%%%%%%%%%%%

\end{document}